\documentclass[aip,jcp,reprint,floatfix,citeautoscript]{revtex4-1}
\usepackage{cmap}
\usepackage[T1]{fontenc}
\usepackage{helvet}

\usepackage{newtxtext}
\usepackage[slantedGreek]{newtxmath}

\usepackage{microtype}
\usepackage[italic,basic]{mathastext}

\usepackage{booktabs}
\usepackage[dvipsnames, table]{xcolor}

\usepackage[pdftex, bookmarks, pdffitwindow=false, pdfstartview=FitH, pdfdisplaydoctitle, colorlinks, plainpages=false, pdftitle={Numerical calculation of free-energy barriers for entangled polymer nucleation},pdfauthor={Xiaoliang Tang, Fucheng Tian, Tingyu Xu, Liangbin Li, Aleks Reinhardt}, pdfpagelabels, hypertexnames, citecolor={cyan},linkcolor={cyan}, urlcolor={violet}, pdflang={en}, hyperfootnotes=false, breaklinks]{hyperref}
\usepackage[all]{hypcap}

\usepackage{graphicx}
\usepackage{amsmath}
\usepackage{flafter}
\usepackage{siunitx}
\usepackage[nodayofweek]{datetime}
\newdateformat{myDate}{\THEDAY\ \monthname[\THEMONTH] \THEYEAR}

\newcommand{\figref}[1]{Fig.~\hyperref[#1]{\ref{#1}}}
\newcommand{\figrefsub}[2]{Fig.~\hyperref[#1]{\ref{#1}(#2)}}

\begin{document}

\title{Numerical calculation of free-energy barriers for entangled polymer nucleation}
\author{Xiaoliang Tang}
\author{Fucheng Tian}
\author{Tingyu Xu}
\author{Liangbin Li}\email{lbli@ustc.edu.cn}
\affiliation{National Synchrotron Radiation Lab, CAS Key Laboratory of Soft Matter Chemistry, Anhui Provincial Engineering Laboratory of Advanced Functional Polymer Film, University of Science and Technology of China, Hefei 230026, China}
\author{Aleks Reinhardt}\email{ar732@cam.ac.uk}
\affiliation{Department of Chemistry, University of Cambridge, Lensfield Road, Cambridge, CB2 1EW, United Kingdom}

\date{\myDate\today}

\raggedbottom

\begin{abstract}
The crystallisation of entangled polymers from their melt is investigated using computer simulation with a coarse-grained model.
Using hybrid Monte Carlo simulations enables us to probe the behaviour of long polymer chains.
We identify solid-like beads with a centrosymmetry local order parameter and compute the nucleation free-energy barrier at relatively high supercooling with adaptive-bias windowed umbrella sampling.
Our results demonstrate that the critical nucleus sizes and the heights of free-energy barriers do not significantly depend on the molecular weight of the polymer; however, the nucleation rate decreases with increasing molecular weight.
Moreover, an analysis of the composition of the critical nucleus suggests that intramolecular growth of the nucleated cluster does not contribute significantly to crystallisation for this system.
\end{abstract}

\maketitle

\section{Introduction}

As the first step of the crystallisation process of polymers, nucleation can often determine the final morphologies of polymer materials; moreover, it can affect mechanical, electrical and optical properties of systems~\cite{Bin2006,Luo2009a,Pei2010,Wang2008}.
Both theoretical and experimental investigations into nucleation behaviour started over a century ago~\cite{Gibbs1873, *Ostwald1897, *Volmer1926, *Becker1935}.
Due to its simplicity, classical nucleation theory (CNT) continues to be widely applied to explain the nucleation process qualitatively~\cite{Oxtoby1998,Karthika2016,*Ford2004,*Anwar2011b},
even though it is well established that deviations from its predictions of the nucleation rate are widespread in many systems~\cite{Oxtoby1998, Merikanto2007, *Courtney1961, *Ruckenstein2005}, particularly in cases where the nucleation free-energy barrier is multi-stage or structured~\cite{Sear2012, *Jacobs2015, *Jacobs2015b}.

Nucleation is a rare event, and as such it can be difficult to investigate both experimentally and computationally.
The probability of a nucleation event occurring depends on the height of the free-energy barrier to nucleation, which can be investigated in computer simulations using rare-event methods such as umbrella sampling~\cite{Torrie1977,Kaestner2011}, metadynamics~\cite{Laio2008} or forward-flux sampling~\cite{Allen2006,*Allen2006b,*Valeriani2007b}.
Barrier crossings have been investigated for a range of systems, from colloids to proteins, ionic crystals and water, and such simulation studies have provided useful insights into the nucleation process especially at the spatial and temporal resolutions which present particular challenges to experiment~\cite{Smallenburg2014,*Valsson2016,*Dickson2010}.

In the context of polymer crystallisation from the melt, Hu \textit{et al.}\ investigated single-chain nucleation using a lattice Monte Carlo simulation and showed that although the chain length determined the free-energy barrier to melting, the free-energy barrier to nucleation was insensitive to it~\cite{Hu2003}.
Yi \textit{et al.}\ studied the nucleation free-energy barrier of a range of polymer melts as a function of the degree of supercooling~\cite{Yi2009,Yi2011,Yi2013}; they showed that for relatively short polymers, interfacial free-energy densities are largely temperature independent~\cite{Yi2011}, suggesting that the nucleation barrier is largely enthalpic in origin, whilst for longer chains, the converse holds, perhaps indicating that chain folding and looping that are possible with longer chains result in complex entropic contributions to the free-energy barrier~\cite{Yi2013}.
Muthukumar and co-workers investigated the role of the entropy of folding of long molecules on the free-energy barrier to nucleation, and showed that for ring polymers, there is a significant free-energy barrier to secondary nucleation that is not present for linear polymers~\cite{Welch2001,Iyer2018, Iyer2019}.
In each of these studies, a different order parameter was used to track and drive the nucleation process.
A unified order parameter in polymer systems is difficult to attain, perhaps in part because structures form hierarchically, molecular chains are often very flexible, and numerous intermediate states exist along the crystallisation pathway.
Such problems become progressively worse as the length of polymer chains increases, making simulations of polymer systems considerably more difficult.
Simulations are also impeded by the very slow dynamics that arise from the inter-connectedness of polymer chains.
Often, special types of simulation moves are used to improve sampling efficiency~\cite{Siepmann1992, *Consta1999, *Cumberworth2018}.
The relaxation time of long-chain polymers increases exponentially with molecular weight~\cite{Farrell1980,Jackson1994}, and thus long, expensive simulations are necessary to attain equilibrium.
The construction of nucleation free-energy profiles of long-chain polymers therefore remains a challenging task using molecular simulations.

It has been shown that in polyethylene melts, the nucleation rate $J$ decreases with increasing molecular weight (MW) and obeys the power law $J \propto M_{\text{n}}^{-H}$, where $M_{\text{n}}^{\vphantom{H}}$ is the number-averaged MW and the parameter $H>0$ is related to the morphology of the crystal~\cite{Ghosh2001}.
By contrast,  in poly(ethylene succinate) samples, $J$ is observed initially to decrease with MW and subsequently to increase beyond some critical value~\cite{Umemoto2003},
suggesting that nucleation may change from an intermolecular regime to an intramolecular one, where cluster growth results from further attachment of segments of molecules whose other segments are already part of the cluster.
Despite the complex interplay of initial nucleation, intramolecular nucleation and entanglement~\cite{Nishi1999}, the height of the free-energy barrier to initial nucleation does not appear to vary with chain length~\cite{Ghosh2001, Nishi1999}.

In this work, we use umbrella sampling within the framework of hybrid Monte Carlo simulations to compute the free-energy barrier to primary nucleation of a polymer and investigate the effect of the molecular weight on the nucleation process and the nucleation rate. By investigating the structure of the critical nucleus, we suggest that for the chain lengths considered here, there is no transition from intramolecular nucleation to intermolecular nucleation with increasing molecular weight.

\section{Model and simulation details}

\subsection{Simulation models}

Polymer crystals are polymorphic; for instance, there exist $\upalpha$, $\upbeta$ and $\upgamma$ forms of polypropylene crystals. The polymorphism remains a challenging task when studying polymer crystallisation using MD simulations, as it typically requires expensive all-atom simulations.
As we are not at this stage investigating the fine structure of the polymer crystal, but are instead interested in generic behaviour of relatively long polymer chains, we use a variant of a widely used coarse-grained poly(vinyl alcohol) (PVA) model~\cite{Reith2001, Meyer2001,*Meyer2002,Luo2009b, Luo2009, Sommer2010, Luo2014} to study the free-energy landscape of nucleation, and thus reduce the computational expense.\footnote{Even though PVA has both monoclinic and orthorhombic polymorphs~\cite{Bunn1948,*Colvin1974}, the coarse-graining results in a hexagonal crystalline polymorph.}
This PVA potential was parameterised by coarse-graining atomistic PVA models and from experimental results~\cite{Reith2001,Meyer2001,*Meyer2002}. We use reduced units throughout, so that $r^\ast=r/\sigma$, $U^\ast=U/\varepsilon$ and $T^\ast = k_\text{B}T/\varepsilon$. In the original parameterisation, these can be converted into real units by using $\sigma = \SI{0.52}{\nano\metre}$, corresponding roughly to the chain diameter of PVA,  and mapping $T^\ast=1$ to a real temperature of $T=\SI{550}{\kelvin}$~\cite{Luo2009b}.
Non-bonded interactions are approximated by a Lennard-Jones (LJ) 9-6 potential~\cite{Meyer2001,*Meyer2002},
\begin{equation}\label{eq:1}
U_{\text{non-bond}}^\ast(r)=1.5114 \left[\left(\frac{\sigma_{0}}{r^\ast}\right)^{9}-\left(\frac{\sigma_{0}}{r^\ast}\right)^{6}\right],
\end{equation}
where $r^\ast$ is the interparticle distance and $\sigma_{0}= 0.89$.
Adjacent beads in a polymer chain, each corresponding roughly to a PVA monomeric unit~\cite{Meyer2001,*Meyer2002}, are bonded with a harmonic potential
\begin{equation}\label{eq:2}
U_{\text{bond}}^\ast(r)=\frac{1}{2} k_{\text{bond}}\left(r^\ast-b_{0}\right)^{2},
\end{equation}
where $k_{\text{bond}} = 2704$ and $b_0 = 0.5$, where these parameters are again derived from comparison to atomistic simulations of PVA~\cite{Luo2009b}.
Finally, the bending of the polymer is described by a tabulated angular potential (see Supporting Data) based on the original parameterisation~\cite{Reith2001,Meyer2001,*Meyer2002, Luo2009b}.

In the coarse-grained PVA potential, the LJ potential is cut and shifted to zero at the minimum of the potential, $r_\text{min}^\ast=1.02$.
The potential is therefore completely repulsive~\cite{Meyer2001,*Meyer2002, Luo2009b}, which improved its computational efficiency.
Effective attractions can then qualitatively be tuned by increasing the pressure of the system.
The angular part of the potential has attractive wells and so the polymer chain becomes stiffer as the temperature is decreased~\cite{Meyer2001,*Meyer2002, Triandafilidi2016}, which in turn can lead to an Onsager-rod-like entropy-driven crystallisation of largely parallel chains at lower temperatures.
However, crystallisation can be, and indeed usually is, driven by attractive interactions between monomers.
To account for this behaviour, we have therefore modified the PVA potential by increasing the cutoff to $r_\text{cutoff}^\ast=1.5$, thus allowing favourable non-bonded interactions with the first neighbour shell of monomers.
We stress that, whilst the results that we obtain using a generic coarse-grained model can provide general insights into the crystallisation of polymer melts, our findings are not expected to reproduce the behaviour of any specific polymer system, not even PVA.
Nevertheless, we have verified that at sensible temperatures and pressures, the local structure of this modified potential is similar to that of the original PVA potential for both the melt and the crystalline states.

\subsection{Hybrid Monte Carlo simulations}
We use a combination of brute-force molecular dynamics (MD) simulations that allow us to probe the natural dynamics of systems on the one hand, and hybrid Monte Carlo (HMC)\cite{Duane1987,*Heermann1990,*Mehlig1992} simulations to probe thermodynamic properties on the other.
Monte Carlo simulations~\cite{Metropolis1953} directly sample the statistical ensemble of choice, and are thus very convenient for probing thermodynamics.
When only local moves are used, such simulations can also yield dynamic information~\cite{Huitema1999}.
However, the efficiency of single-particle moves can rapidly decrease when collective motion becomes important~\cite{Reinhardt2012b}.
Since the relaxation time of polymer molecules increases exponentially with molecular weight, the longer the polymer chains we wish to simulate, the more important collective molecular motion becomes.
We therefore use the HMC scheme, where short MD simulations are used instead of the usual single-particle trial moves within an overarching Monte Carlo simulation.
These short MD simulations must be time-reversible and symplectic to obey detailed balance~\cite{Duane1987,*Heermann1990,*Mehlig1992}.
We therefore use MD simulations in the microcanonical ensemble with a velocity Verlet time integrator, and we assign initial velocities to particles from a normal distribution with zero mean and a variance of $k_{\text{B}} T/m$ to satisfy the Maxwell--Boltzmann distribution at the temperature of interest.
We use an in-house code to perform all calculations, except that molecular dynamics simulations, both brute-force and within HMC, are performed using \textsc{Lammps}~\cite{Plimpton1995}.
In HMC simulations, \textsc{Lammps} is interfaced as a Python library.

We use a time step of $\updelta t = 0.001\tau$ and $\updelta t = 0.01\tau$ in HMC and brute-force MD simulations, respectively, where $\tau=(\sigma^2 m/\varepsilon)^{1/2}$ is the effective unit of time.
The mass $m$ does not generally correspond to the molecular mass of each bead, since coarse-graining removes some degrees of freedom which can slow a system's dynamics, resulting in a larger effective value for $m$ when mapping back to real time.
For the original PVA model, $\tau\approx \SI{3}{\pico\second}$~\cite{Sommer2010}, which can give an indication of the typical relevant timescales; however, we report our results in terms of $\tau$ for generality.

The polymer chain lengths range from $l=20$ to $l=300$ across simulations. The entanglement length, $N_{\text{e}}$, of the original PVA model is $\sim$30 monomers~\cite{Luo2013}, and so the chain lengths in this work cover both unentangled and entangled states. We use approximately 10000 polymer beads in each simulation, but the systems have different numbers of chains, namely $\lfloor 10000/l \rfloor$, where $\lfloor \cdot \rfloor$ denotes the floor function. In order to verify that finite-size effects are not significantly affecting our results, we have confirmed that a simulation of 5000 polymer beads with a chain length of $l=100$ results in essentially identical behaviour (see Supporting Data), indicating that the systems are sufficiently large to probe primary nucleation under the conditions studied.

In HMC simulations, in which only the MC simulation is subjected to a biassing potential (see Subsect.~\ref{subsec-US}), we typically use 10 MD steps for each MC move so as to achieve a $\sim$60\,\% acceptance rate. In order to minimise finite-size effects on nucleation~\cite{Wedekind2006}, we perform these simulations in an isobaric ensemble at $P=2\sigma^{3}/\varepsilon$, which we implement by scaling the box using trial moves in $\ln(V)$~\cite{Eppenga1984, Frenkel2002}.
In brute-force MD simulations, we implement the isothermal-isobaric ensemble with a Nos\'{e}--Hoover thermostat\cite{Nose1984, *Hoover1985} and a Parrinello--Rahman-like barostat~\cite{Parrinello1981, *Martyna1994, *Shinoda2004}.
Sample input scripts detailing all parameters used are provided in the Supporting Data.

\begin{figure}[tbp]
\centering \includegraphics[width=8cm]{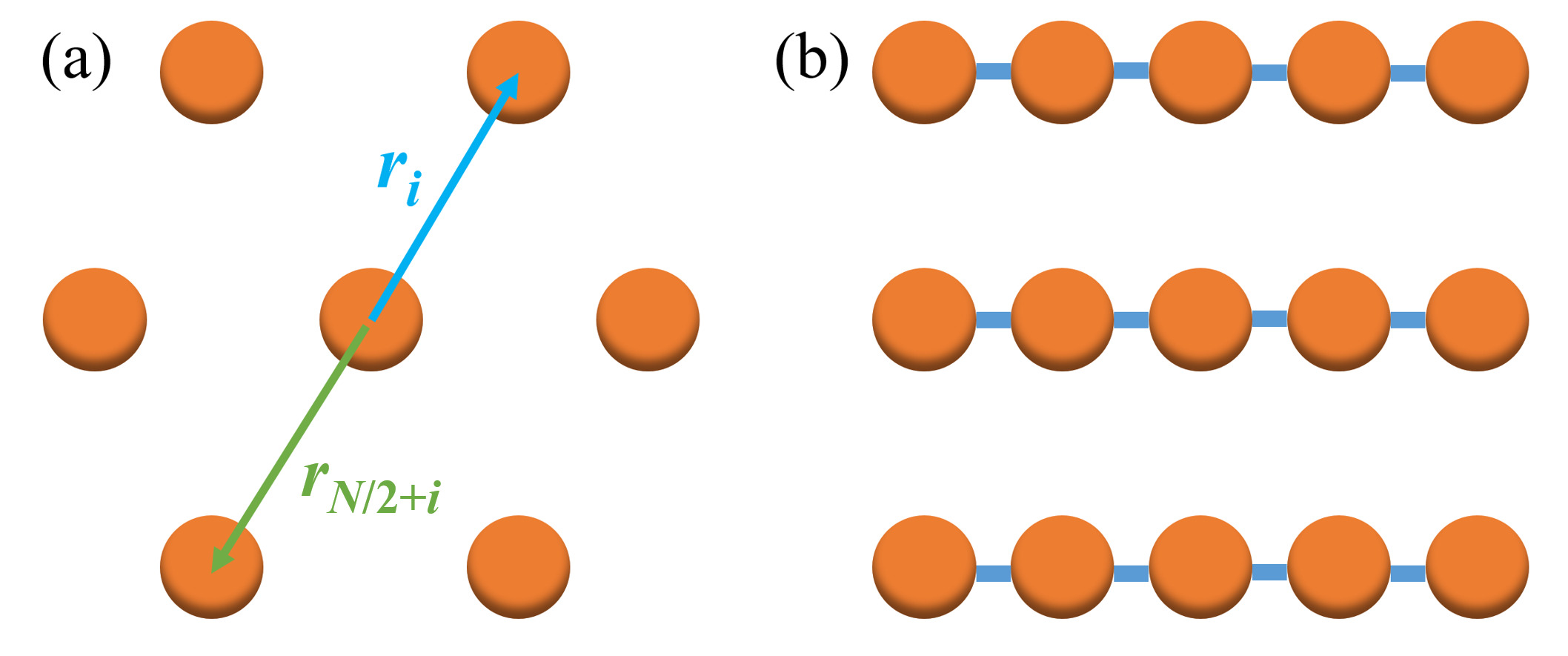}
\caption{A schematic illustration of the (a)~top and (b)~side view of the crystalline structure of a polymer using the coarse-grained PVA model. Orange circles represent beads in the polymer, and thick blue lines indicate bonds between the beads. One pair of the bond vectors used in the order parameter calculation is illustrated in panel~(a).}\label{fig:CGPVAviews}
\end{figure}

\subsection{Local order parameter}
In nucleation studies, a significant challenge is the identification of particles in the two phases of interest.
In previous simulation studies, several distinct order parameters were used to track and drive the nucleation process: the total number of disordered monomer units~\cite{Hu2003}; the largest number of neighbouring chains with the same orientation~\cite{Yi2009,Yi2011,Yi2013}; and the lamellar thickness~\cite{Welch2001,Iyer2018}.
To enable a comparison to classical nucleation theory, a convenient order parameter to use is the size of the largest crystalline cluster; however, such an order parameter can only be computed if solid particles have been correctly identified.
Since `phases' are macroscopic concepts, identifying individual particles as being solid-like or liquid-like at the microscopic scale is fraught with difficulties.
One of the most commonly used approaches of achieving such an identification for spherical particles is the Steinhardt--Ten Wolde order parameter~\cite{Steinhardt1983,TenWolde1996}, which has successfully been used both in simulations and in experiment~\cite{Tan2014}.

\begin{figure*}
\centering \includegraphics[width=16cm]{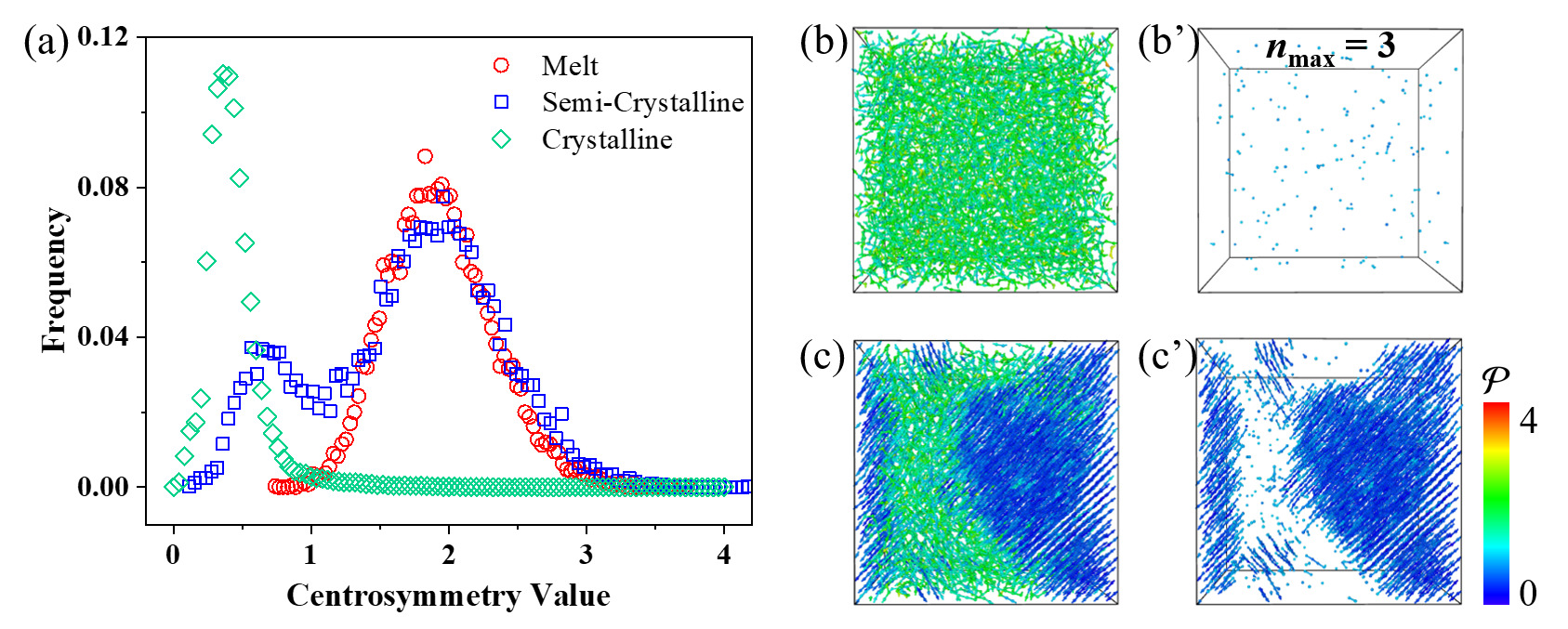}
\caption{(a) Distributions of centrosymmetry values $\mathcal{P}$ of the melt (red circles), semi-crystalline (blue squares) and crystalline (green diamonds) states with a histogram bin width of 0.04. The snapshots extracted from brute-force MD trajectories (b) and (c) are coloured by centrosymmetry values. The corresponding solid-like particles only are shown in (b') and (c'), respectively.} \label{fig:centrosymmetry}
\end{figure*}

There is no well established local order parameter used in the polymer community to achieve cluster classification because polymer molecules are conformationally very flexible, and many polymorphs of polymer crystals exist~\cite{Tang2017,Nicholson2016,Luo2011}.
In the present case, the simplified coarse-grained polymer model exhibits hexagonal symmetry in the plane perpendicular to the direction of backbones (\figrefsub{fig:CGPVAviews}{a}).
Particle identification can therefore be achieved more readily than would be the case with more complex models.
We use the centrosymmetry parameter $\mathcal{P}$ proposed by Hamilton and co-workers~\cite{Kelchner1998},
\begin{equation}
\mathcal{P}=\sum_{i=1}^{N/2}|\boldsymbol{r}_{i}+\boldsymbol{r}_{N / 2}|,
\end{equation}
where $\boldsymbol{r}_{i}$ is the bond vector from the central particle to the $i$th neighbouring particle and $N$ is the number of neighbouring particles.
This parameter divides all the bond vectors into $N/2$ pairs, and the sum across all pairs adds up to zero in a perfect crystal [\figrefsub{fig:CGPVAviews}{a}].
The lower the value of $\mathcal{P}$ is, the more ordered the structure in question is.
The distribution of $\mathcal{P}$ is shown in \figrefsub{fig:centrosymmetry}{a}.
We classify particles whose centrosymmetry value is lower than 1 as solid-like, and melt-like otherwise.
In Fig.~\ref{fig:centrosymmetry}, we show how such phase identification works in practice.
The isolated beads in \figrefsub{fig:centrosymmetry}{b'} indicate that only very small `solid-like' clusters exist in the molten state.
On the other hand, large solid clusters can be observed in the semi-crystalline state.
We consider solid-like beads within a distance of $1.05\sigma$ of one another as belonging to the same crystalline cluster, and the number of beads within one cluster represents the cluster size $n$.
We use the size of the largest such cluster, $n_{\text{max}}$, as our order parameter.
Of course these choices may affect the final nucleation free-energy profile, and we return to this point in Subsection~\ref{subsect:choiceOfRxnCoord}.

\subsection{Umbrella sampling with an adaptive biassing potential}\label{subsec-US}
\begin{figure}[tbp]
\centering \includegraphics[width=8cm]{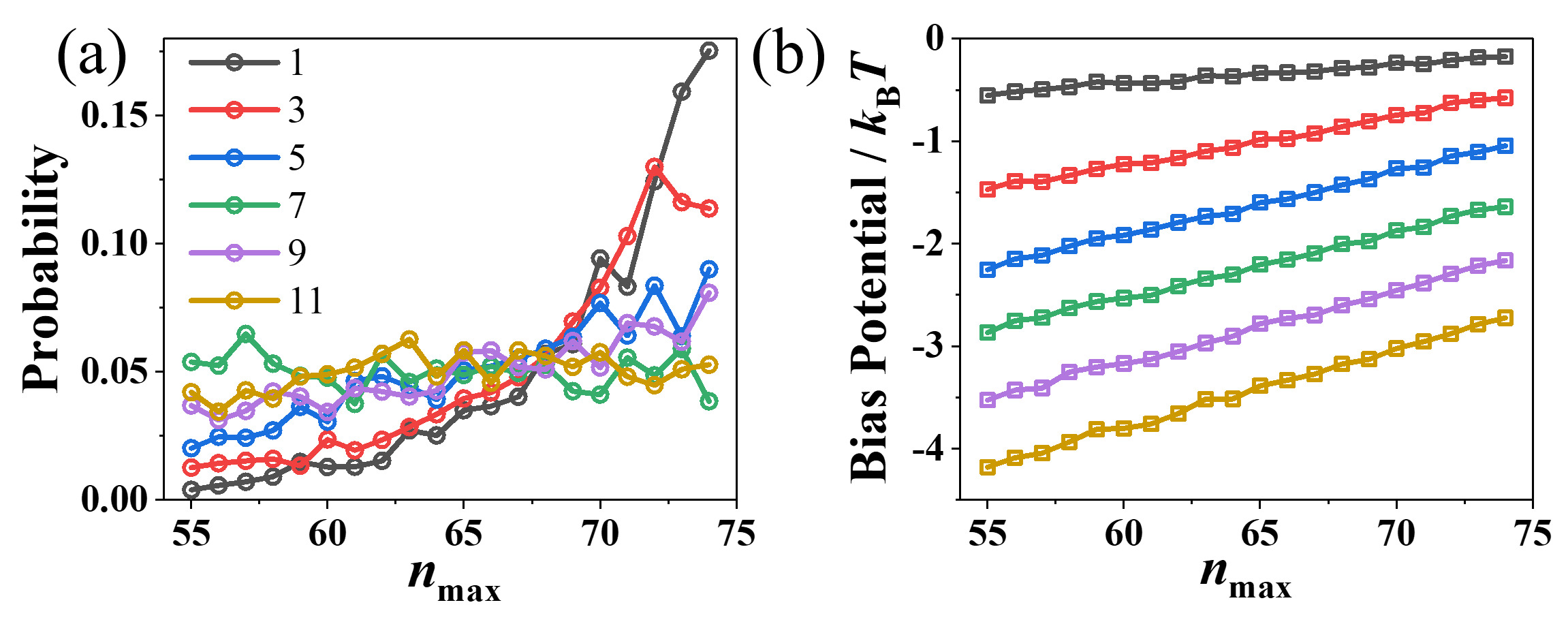}
\caption{Umbrella sampling equilibration. By way of illustration, we show results for a single post-critical window with $n_{\text{max}}\in[55,\,74]$. In (a), we show the frequency distribution of $n_{\text{max}}$ of successive umbrella sampling iterations (as labelled in the legend) within the window. In (b), we show the biassing potential and its variation at the end of each umbrella sampling iteration.
}\label{fig:US-equilibration}
\end{figure}

The free-energy barrier to nucleation is $\upDelta G(n)/k_\text{B}T \equiv -\ln (N_n/N)$, where $N_n$ is the number of clusters of size $n$ and $N$ is the total number of particles.
For small clusters, this free energy can be computed from brute-force MD simulations in the isothermal-isobaric ensemble.
However, for larger cluster sizes, $N_n/N$ rapidly decreases and approaches the probability that the largest cluster in the system is of size $n$~\cite{Reinhardt2014}.
For sufficiently large clusters, we can therefore compute the free energy $G(n_{\text{max}})$ only.
However, such clusters are usually very rare, and to find this free energy, we use the umbrella sampling technique~\cite{Torrie1977}, which allows us to sample regions of phase space with a low sampling probability by introducing a biassing potential.
We split the region of cluster sizes of interest into several partially overlapping windows to speed up equilibration~\cite{Chandler1987, VanDuijneveldt1992,Auer2004}.
In practice, we use eight windows along the order parameter $1 \le n_{\text{max}}<120$; each window overlaps with its neighbours by 4 units.
We compute the free energy within the first window ($n_{\text{max}} \in [1,\,15)$) directly with brute-force MD, and that of the remaining windows with umbrella sampling.

Instead of traditional quadratic biassing potentials, we use adaptive umbrella sampling~\cite{Mezei1987}, where the bias is gradually adapted between simulations to enable the entire window in order-parameter space to be sampled.
The initial biassing potential $U^\text{b}(n_{\text{max}})$ is set to zero.
To update the biassing potential, we combine the biassing potential used in a set of simulations ($U_\text{old}^\text{b}(n_{\text{max}})$) with the frequency distribution of the order parameter ($f(n_{\text{max}})$) to give
\begin{equation}\label{eq:14}
U_\text{new\vphantom{l}}^\text{b}(n_{\text{max}})=U_\text{old}^\text{b}(n_{\text{max}})+ k_{\text{B}} T \ln f(n_{\text{max}}).
\end{equation}
In practice, in the initial stages of a simulation, $k_{\text{B}} T \ln f(n_{\text{max}})$ can fluctuate drastically if the entire region is not properly sampled, and to minimise hysteresis effects, we limit any update of the biassing potential to a maximum of $1k_\text{B}T$ in a single iteration.
The biassing potential is updated until equilibrium is reached, when $f(n_{\text{max}})$ is uniformly distributed in the entire window and so any update to the biassing potential adds only a constant term for all $n_\text{max}$ and is thus no longer meaningful.
The biassing potential is then expected to be the negative of the free energy for each $n_{\text{max}}$.
We show an illustration of how the biassing potential is updated in Fig.~\ref{fig:US-equilibration}.
In our simulations, we run simulations for 12000 MC steps between each successive update to the biassing potential.

Finally, we note that there is no absolute zero for the Gibbs energy, and so the free energies obtained from different umbrella sampling windows can be shifted by a constant from one another.
Since the windows by construction have ranges that overlap with one another, the free energy of one pair of such overlapping points can be matched up across windows.
If the remaining points are also well matched up, this is a useful indicator that the windows are well equilibrated.
More complex procedures, such as the weighted-histogram~\cite{Kumar1992} or multi-Bennett acceptance ratio~\cite{Shirts2008} methods, can alternatively be used.

\section{Results}
\subsection{Free-energy profiles of primary nucleation}
Classical nucleation theory provides a qualitative description of the homogeneous nucleation process, and the relation between crystallisation temperature and the height of the free-energy barrier has been investigated in some detail in the field of polymer crystallisation~\cite{Armistead2002}.
Here, we focus in particular on the effect of molecular weight, or, equivalently, the chain length $l$, of the polymer on its primary nucleation.

\begin{figure}[tbp]
\centering \includegraphics[width=8cm]{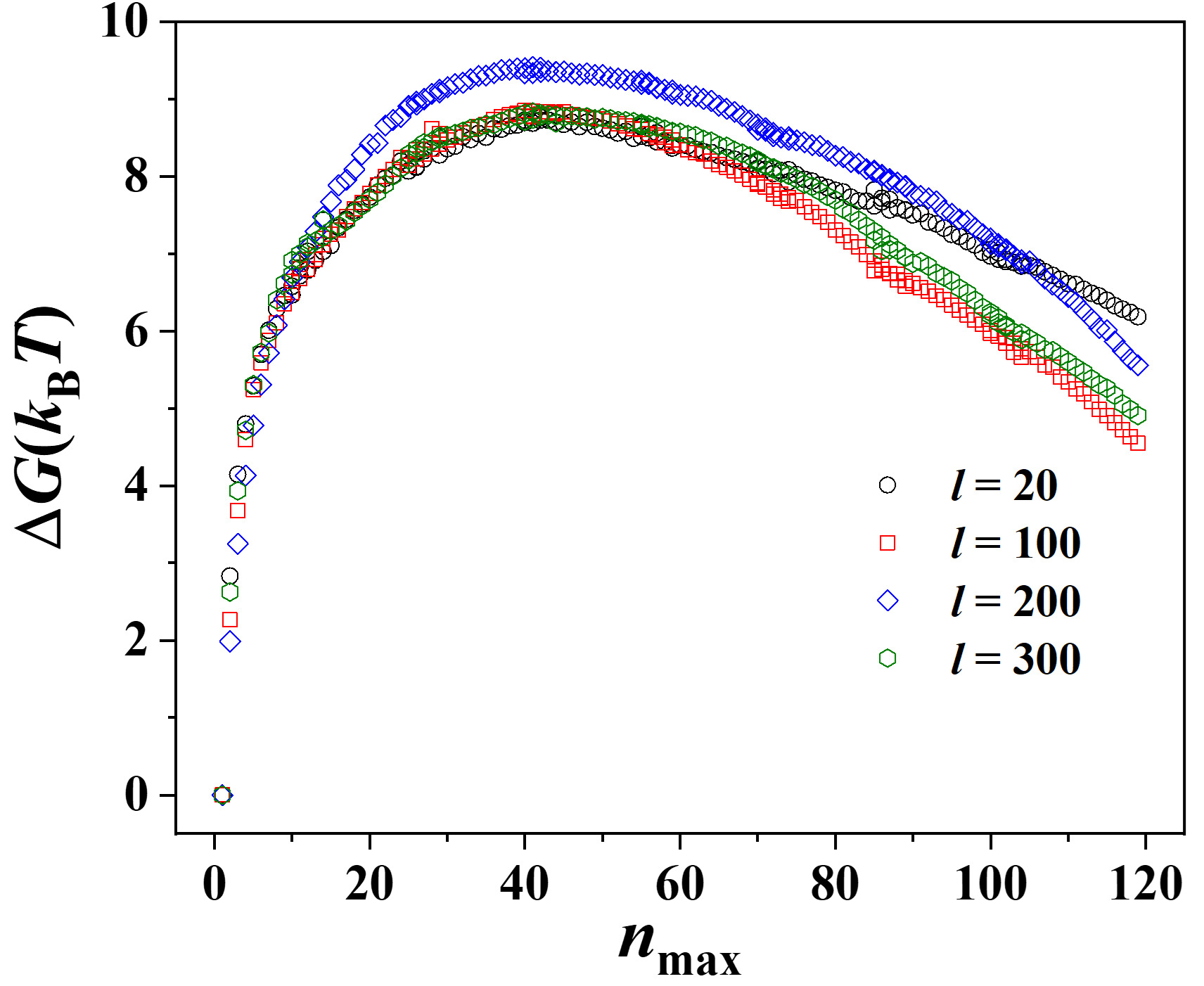}
\caption{Nucleation free-energy profiles of polymers with different chain lengths.}\label{fig:nucBarriers}
\end{figure}

\begin{table}[t!]
\centering
\caption{Simulation temperatures used for different chain lengths.}\label{T:1}
\small
\setlength{\tabcolsep}{0.5em}
\begin{tabular}{cccccccc}
\toprule
Chain length $l$ & 20 & 50 & 100 & 150 & 200 & 250 & 300 \\
Temperature $T^\ast$ & 0.80 & 0.81 & 0.82 & 0.83 & 0.84 & 0.84 & 0.84\\ \bottomrule
\end{tabular}
\end{table}

The equilibrium melting temperature of each system considered is not known, and this temperature generally depends on the chain length.
In order to compare nucleation behaviour under similar conditions, we first determine the hysteresis behaviour of polymer melts as the system is gradually cooled until it forms a (semi)crystalline state, and then subsequently heated until it melts again. The temperatures at which freezing and subsequent melting occur bracket the thermodynamic melting point of the system.
The temperature at which spontaneous crystallisation is observed increases with chain length, and so, in order to compare nucleation at a similar degree of supercooling, we study the nucleation behaviour at slightly different temperatures for each chain length, as listed in Table~\ref{T:1}.
The temperatures listed in the table are estimates based on the behaviour of each system in brute-force simulations rather than the true thermodynamic melting temperature, which is not known.
However, although the details of the nucleation behaviour can change as a function of  supercooling, the supercooling in each case is large and so the qualitative behaviour is unlikely to be significantly affected by the precise simulation temperature chosen.

The temperatures chosen in this way result in similar free-energy barriers to nucleation, as shown in \figref{fig:nucBarriers} for different chain lengths;
the heights of the free-energy barriers are all fairly close to $\sim${}$9k_{\text{B}} T$,
with the relatively low barrier heights not unexpected because the simulations are performed at high supercooling.
Interestingly, the size of the critical nucleus ($n^{*}$) is close to 40 across all chain lengths.
To benchmark this result, we also performed  brute-force MD simulations starting from the critical nuclei ($n_{\text{max}} \approx 40$) of each system. Of the forty simulations performed for each system, roughly half resulted in cluster growth and the other half in shrinkage, demonstrating that the critical cluster was indeed obtained.

\begin{figure}[tbp]
\centering \includegraphics*[width=8cm]{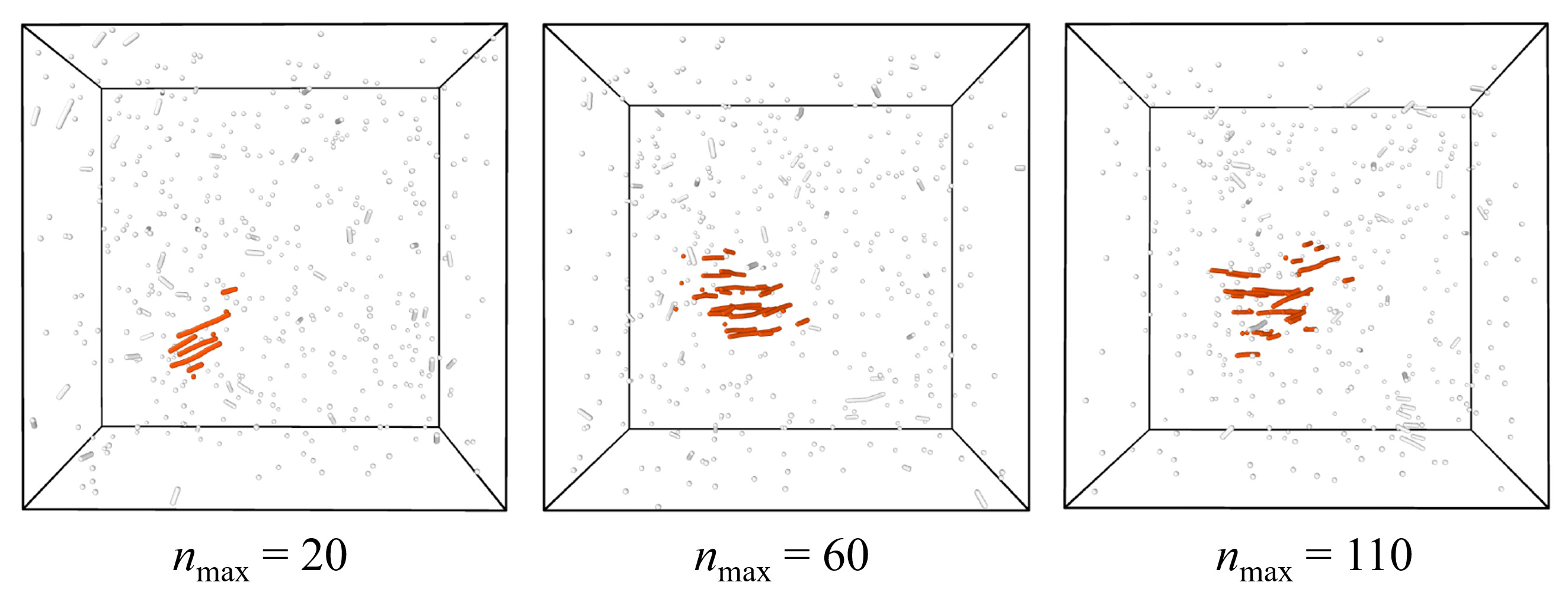}
\caption{Snapshots of the $l=20$ system during nucleation, when the largest cluster in the system comprised 20, 60 and 110 beads. Grey beads are solid-like particles with a centrosymmetry value less than 1. These beads are connected if they are adjacent within the same chain. Red beads and bonds show the largest crystalline cluster in the system, which usually appears to be largely ellipsoidal.}\label{fig:snapshots}
\end{figure}

Since the coarse-grained potential we use is largely governed by excluded volume interactions, the potential energy does not significantly change during the primary nucleation process, indicating that the nucleation free-energy barrier arises from a disfavourable change in entropy~\cite{Reinhardt2013c}.
If we assume, to a first order of approximation, that classical nucleation theory applies in this case, we can estimate an approximate interfacial free-energy density by using a non-linear least-squares fit of the free-energy profile of \figref{fig:nucBarriers} to the polynomial $\beta\upDelta G(n_{\text{max}}) = a_{0}n_{\text{max}} + a_{1}n_{\text{max}}^{2/3}+ a_{2}$ to account for the bulk and surface contributions to the free energy, where $a_2$ shifts the origin of the curve.
For the $l=100$ case, the fitting parameters were determined to be $a_{0} = -0.4$, $a_{1} = 2.0$ and $a_{2} = 0.5$.
If we further hypothesise that the nucleus is spherical -- which is not an unreasonable first-order estimate (see Fig.~\ref{fig:snapshots}), but which is unlikely to be very accurate for polymer nucleation\cite{Yi2011} --, the isotropic interfacial free-energy density can be estimated as $\gamma = a_{1} k_\text{B}T \times(36\uppi/\rho^{2})^{-1/3}$, where $\rho$ ($\sim${}$ 2.8\sigma^{-3}$) is the number density of the beads in the crystal phase.
In this case, we can estimate the interfacial free-energy density to be $\gamma \approx 0.7\varepsilon\sigma^{-2}$.

\begin{figure}[tb]
\centering \includegraphics[width=8cm]{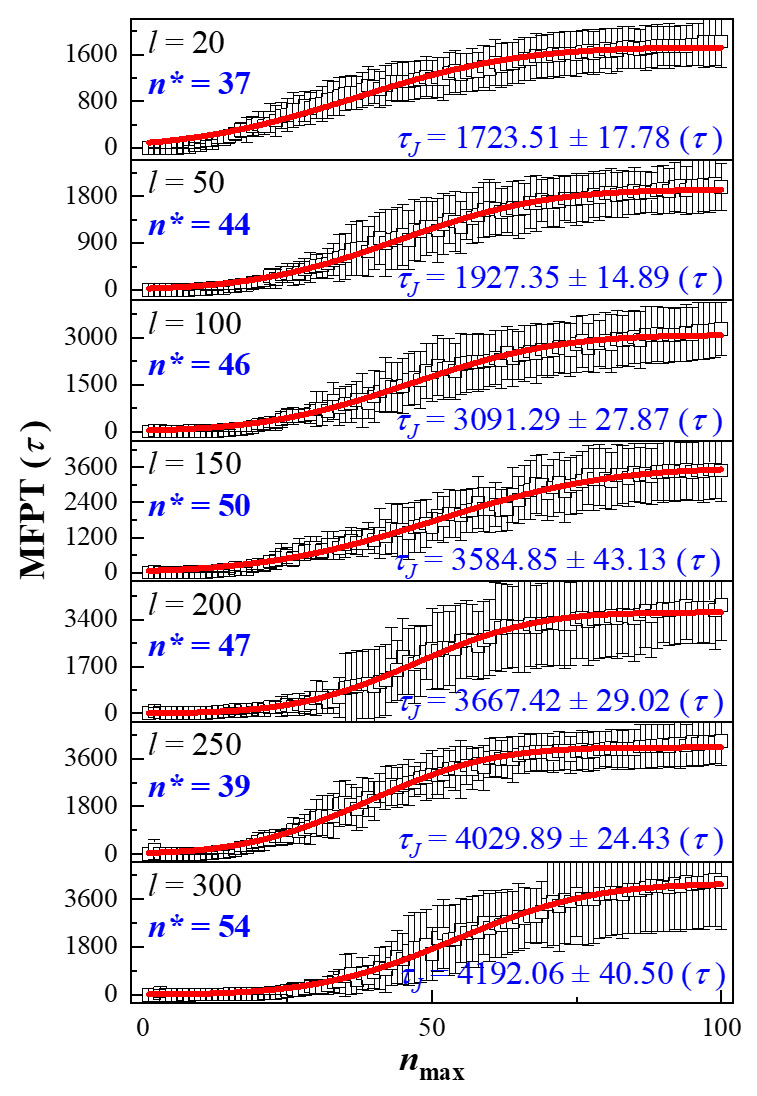}
\caption{MFPT of systems with different chain lengths. The critical nucleus size fluctuates around 40, consistent with the HMC umbrella sampling results.
Mean first-passage times (in black) are computed from forty independent brute-force MD trajectories in each case. Error bars give the standard error of the mean.
We estimate $n^\ast$ by fitting these times to Eq.~\eqref{eq:MFPT}, with $\tau_{J}$, $n^\ast$ and $c$ being free fitting parameters, and we plot the fitted curve in red.
}\label{fig:MFPT}
\end{figure}

\subsection{Comparison with brute-force MD}

Using HMC with umbrella sampling enables us to compute the free-energy profiles of polymers as a function of their molecular weight.
Although the natural dynamics of the system are significantly affected by coarse-graining, we can also gain further insight into the nucleation process without using a biassing potential in brute-force molecular dynamics simulations.
The mean first-passage time (MFPT)~\cite{Wedekind2007b} is a useful measure of the dynamics of activated processes, including polymer primary nucleation.
Information about the critical size of the nucleus and the rate of nucleation can be extracted directly from the trajectories of brute-force MD simulations using this method, provided that these rare events are accessible in MD simulations.
In the light of the high supercooling considered here, nucleation events can indeed be observed in brute-force MD simulations, and the results obtained by the two approaches can be compared directly.

Forty individual brute-force MD simulations were run for each polymer length considered over a total simulation time is $10^{6}$ MD steps, during which some of the trajectories remain in the molten state whilst others crystallise.
We measure the MFPT from these nucleation events by fitting to~\cite{Wedekind2007b}
\begin{equation}\label{eq:MFPT}
\tau(n_{\text{max}})=\frac{\tau_{J}}{2}\{1+\operatorname{erf}[(n-n^{*})c]\},
\end{equation}
where $\tau(n_\text{max})$ is the MFPT of every $n_{\text{max}}$, $\tau_{J} =1/JV$ is a parameter associated with the nucleation rate $J$ and volume $V$, $n^{*}$ is the critical nucleus size, and $c$ is the scaled Zeldovich factor~\cite{Zeldovich1943, Auer2001}, $Z \equiv \sqrt{|G^{''}(n^{*})|/2\uppi k_{\text{B}} T} =c/\sqrt{\uppi}$.

In Fig.~\ref{fig:MFPT}, we show the MFPT as a function of cluster size for a range of chain lengths.
There is a considerable standard error of the MFPT; such a large error is the result of limited sampling points, as only forty individual brute-force MD are performed for each system, and only those which resulted in successful nucleation are considered in the MFPT calculation~\cite{Yi2009}.
Moreover, as the free-energy barrier is not very high ($\sim${}$9k_{\text{B}}T$), it is not straightforward to separate the nucleation and growth processes precisely~\cite{Wedekind2007b}.
Nevertheless, Fig.~\ref{fig:MFPT} shows that $n^{*}$ fluctuates around 40 in all brute-force MD simulations, which is consistent with the results of hybrid MC simulations with umbrella sampling.
Interestingly, $\tau_{J}$ increases with increasing chain length, which suggests that the rate of nucleation will decrease with molecular weight; we discuss this in more detail below.

\subsection{Choice of reaction co-ordinate}\label{subsect:choiceOfRxnCoord}

The free-energy profile is reconstructed as a function of a particular order parameter.
The free energy computed is effectively a projection of the potential energy landscape onto this order parameter; however, there is no guarantee that the chosen order parameter is in fact the true reaction co-ordinate.
Choosing a different order parameter can drastically affect the free-energy profile~\cite{Frenkel2013}, and so the choice of a suitable parameter is of particular importance.

When calculating the order parameter $n_{\text{max}}$ in this work, we used two cutoffs in our definitions of clusters.
Namely, we identified solid-like particles as those with $\mathcal{P} < 1$, and we used a neighbour cutoff of $r_\text{cut} = 1.05 \sigma$ in cluster analysis.
It has been shown that although choosing a more conservative cutoff results in a smaller critical cluster size, when chosen within reason, it does not significantly affect the rate of nucleation~\cite{Wedekind2007b}.
However, the choice of cutoff for the centrosymmetry parameter may affect the results more significantly.
A smaller value of $\mathcal{P}$ corresponds to a more ordered structure; however, in the initial stages of primary nucleation, and particularly so in polymer systems, the conformational flexibility of molecules precludes very significant positional ordering~\cite{Tang2017,Tang2019,Martins2013}.The height of the free-energy barrier may therefore be overestimated if too small a cutoff is used.

Based on the distribution of centrosymmetry values (\figrefsub{fig:centrosymmetry}{a}), we used $\mathcal{P}<1$ as the criterion to classify particles.
We show several snapshots of the nucleation process of the $l=20$ system in Fig.~\ref{fig:snapshots} using this criterion.
Many isolated beads exist throughout the nucleation process, and several short molecular segments are embedded in the nucleus.
As the centrosymmetry criterion is not overly strict in particle classification, the free-energy barrier would be underestimated if the umbrella sampling were performed in the first window.
Therefore, all the clusters are considered in the first window and the free-energy difference is calculated from the distribution of cluster sizes.

In general, defining a local order parameter is a challenging task in polymer systems, especially in all-atom simulations, and the polymorphism of polymer crystals makes the task more difficult still.
A single order parameter can hide information from orthogonal dimensions.
Given that intermediate states may exist in such orthogonal dimensions, it may be prudent to consider multiple physically reasonable order parameters and map out a free-energy landscape.
However, in the case of the coarse-grained potential we use here, the good agreement in the size of the critical cluster between brute-force MD simulations and the free-energy calculation gives us a degree of confidence in the robustness of the order parameter used.

\section{Discussion}
\begin{figure}[tbp]
\centering \includegraphics[width=8cm]{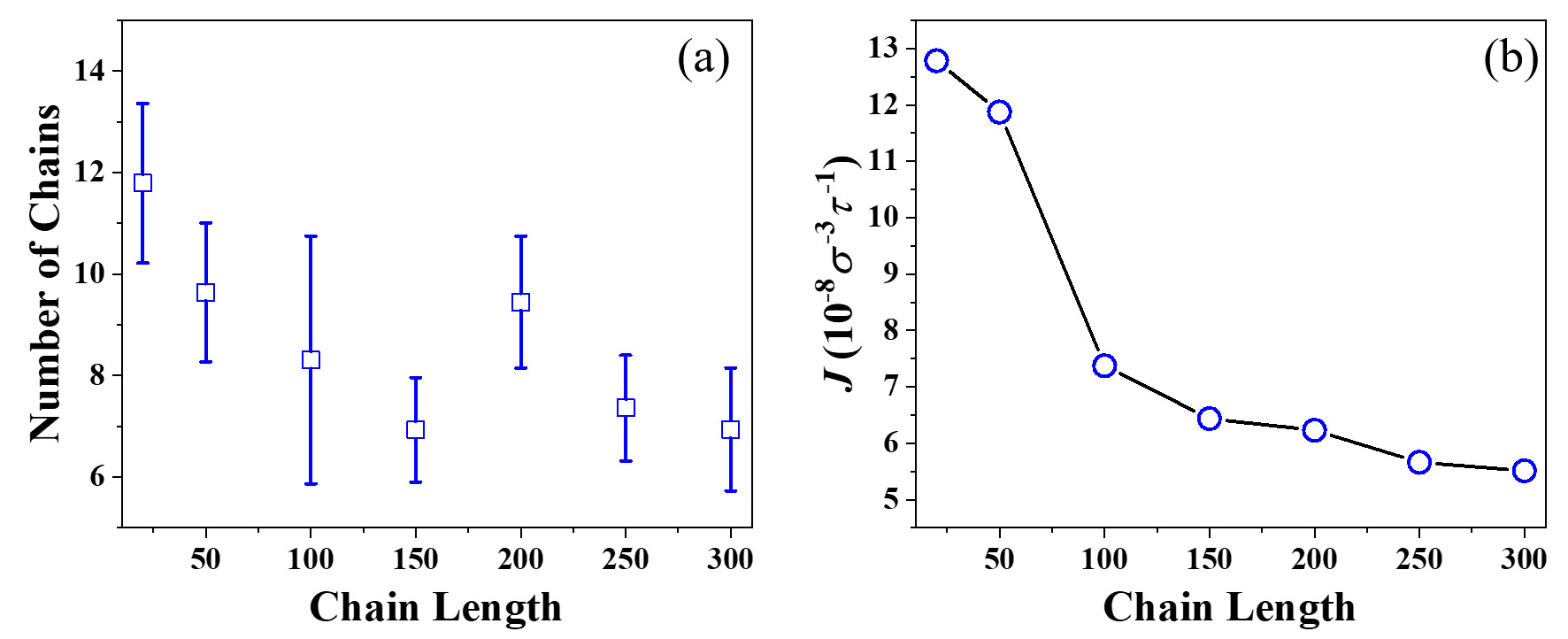}
\caption{(a) The number of different chains within the critical nucleus as a function of chain length, averaged over 20 independent configurations of the critical nucleus in each case. Error bars give the standard deviation.
(b) The nucleation rate as the function of chain length. These data are calculated using the fitting results of Fig.~\ref{fig:MFPT}.}\label{fig:nucRate}
\end{figure}
\subsection{Effect of molecular weight on primary nucleation}

Both HMC and brute-force MD simulation results indicate that neither the height of the free-energy barrier nor the size of the critical nucleus for primary nucleation change significantly with chain length (or, equivalently, with molecular weight), which is consistent with previous experimental work~\cite{Ghosh2002,Umemoto2003}, in which classical nucleation theory was used to estimate these parameters.
One possible interpretation of this is that only some sections of individual chains form the critical cluster, and so the remainder of the chain does not play an important role in the nucleation of the critical cluster.
To test whether this explanation is borne out in our simulations, we analysed the composition of the critical nucleus.
The critical nucleus comprises approximately 40 particles, but in \figrefsub{fig:nucRate}{a} we show that the average number of distinct molecular chains embedded in the critical nucleus for each system is relatively small and at least initially decreases with increasing chain length.
The average number of segments that are embedded in the critical nucleus is $\sim$7, which for most systems considered is very short compared to the overall chain length.
Since we keep the average density of the monomeric particles of the systems constant, increasing the chain length of the polymers does not lead to any significant additional crowding, and so it is not unreasonable that the free-energy barrier should be largely independent of the chain length.
Moreover, the fact that for sufficiently long chains, neither the size of the critical nucleus nor the number of distinct chains in the critical nucleus significantly change strongly suggests that there is no changeover from inter- to intramolecular nucleation as the chain length increases for this system for the chain lengths considered. The free-energy barrier appears to correspond to a largely intermolecular nucleation pathway.

\begin{figure}[tbp]
\centering \includegraphics[width=8cm]{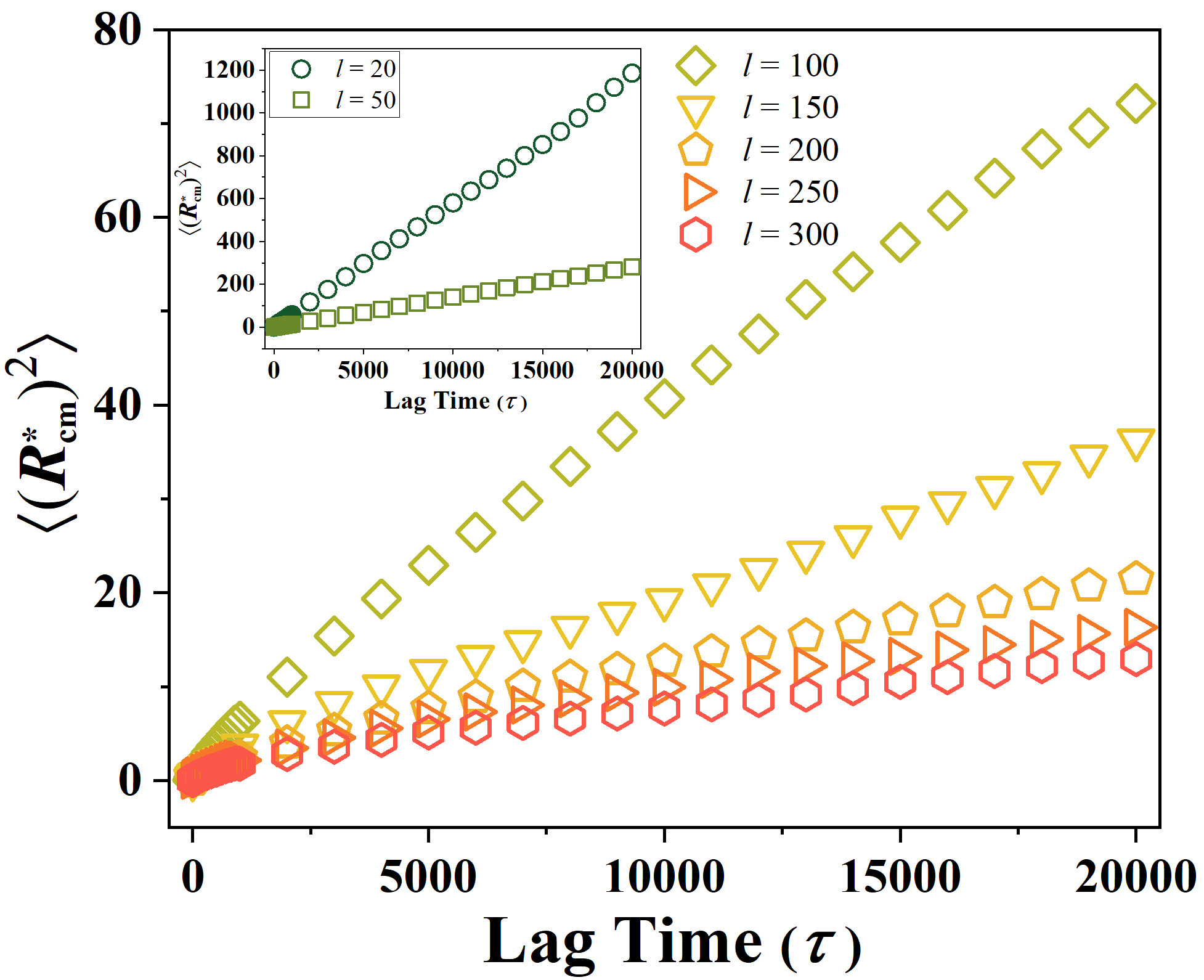}
\caption{Time-averaged mean squared displacement of the centre of mass $\langle (R^\ast_\text{cm})^2 \rangle$ as a function of time and chain length, averaged over 10 chains for each system. The inset shows data for $l = 20$ and $l = 50$.
}\label{fig:MSD}
\end{figure}

How the MW affects the nucleation rate $J$ is of particular interest given the unusual behaviour observed in previous experimental work~\cite{Ghosh2001, Umemoto2003}.
In \figrefsub{fig:nucRate}{b}, we show the nucleation rate $J=1/V\tau_J$ for our systems as a function of chain length, computed from the fitting parameter $\tau_J$ used in the MFPT analysis and the volume of each system.
As the chain length increases, the nucleation rate seems broadly to decrease, and there does not seem to be any turning point in this trend over the range of chain lengths considered.
Empirically, the nucleation rate is often described by the relation $J = J_0 \exp(-\upDelta G^*/k_{\text{B}}T)$, where $J_0$ is a parameter related to diffusion.
In polymer melts, the motion of a chain is restricted by neighbouring chains, and this `entanglement' restriction becomes progressively stronger as the MW increases.
The entanglement length for the original PVA model was determined to be between 30 and 50~\cite{Luo2013}, which appears to lead to a significant slowdown of chain dynamics beyond such chain lengths, consistent with the slowdown seen in \figrefsub{fig:nucRate}{b}.
In order to confirm this conjecture, we plot the time-averaged mean squared displacement of the centre of mass of each polymer chain $\langle (R_\text{cm})^2 \rangle$ for each system in the molten state.
In the inset of Fig.~\ref{fig:MSD}, the $\langle (R_\text{cm})^2 \rangle$ of $l = 20$ and $ l = 50$ exhibits linear relations with lag time, consistent with the Einstein relation $\langle R^2 \rangle = 6Dt$.
We can estimate the diffusion coefficient of these two systems  as approximately $0.0100\,\sigma^2/\tau$ and $0.0017\,\sigma^2/\tau$, respectively.
On the other hand, longer chain systems do not result in a linear relation between $\langle (R_\text{cm})^2 \rangle$ and the lag time, which indicates that their motion is subdiffusive within the timescale of our simulations.
Nevertheless, we may utilise the long-time behaviour of different systems to predict their mobilities.
The mean squared displacement at the final timestep decreases monotonically when the chain length increases from 100 to 300, suggesting that diffusion becomes progressively more difficult when molecular chains are longer.

Therefore even if the height of the nucleation free-energy barrier is roughly constant as a function of $l$ at the same level of supercooling, the decrease in diffusivity can result in a significant reduction in the effective nucleation rates, which can be reflected in differences in the pre-factor $J_0$ in the rate expression.
However, whilst we have demonstrated that the diffusive behaviour of the centre of mass of the polymer chains changes as a function of $l$, the microscopic nature of the effect of slowing diffusion is not obvious, and so quantifying its effect on the nucleation behaviour is difficult.
In particular, there are likely also to be other differences in diffusivity beyond that of the centre of mass. For example, beads closer to the end of a polymer can behave very differently from those at the centre, and with polymer molecules, there are many hierarchical levels of distinct diffusive behaviour as a function of lag time~\cite{Kresse2015} that complicate matters further.
It would be interesting in future work to investigate the statistics of which parts of polymer molecules crystallise first and whether anything significant can be said about the local diffusivity of beads within those parts of the chains both in the melt and within the critical cluster of the growing crystallite.

Moreover, while we can conclude from simulations that for the range of chain lengths considered in this work, the nucleation rates decrease as a function of increasing chain length using our model, by contrast, in experiments on poly(ethylene succinate)~\cite{Umemoto2003}, an increase of nucleation rate with increasing MW for sufficiently long polymers was observed.
The reason for the differing behaviour may simply be that the chain length in our simulations is not sufficiently long. However, such behaviour might also originate from the molecular structure, where oxygen atoms in the backbones may result in hydrogen bonding, which may significantly affect the diffusion behaviour of the polymer chains.
Indeed, although hydrogen bonding also exists in PVA systems, interactions between molecules cannot be accurately reproduced with the generic coarse-grained model we are using. 
Moreover, when crystallising, polymer chains naturally fold, and crystallised samples often exhibit significant adjacent re-entry packing~\cite{Wang2019,*Hong2015}.
Although we did not observe intramolecular primary nucleation, this of course does not mean that subsequent polymer folding cannot occur during crystal growth.
The precise mechanism of the putative transition from inter- to intramolecular nucleation and the effect of chain length on the subsequent growth therefore remain unclear and deserve further investigation.

\section{Conclusion}
We performed a set of hybrid MC simulations to study the effect of chain length on the primary nucleation of chain polymers from the melt, and used umbrella sampling to compute free-energy profiles using the size of the largest crystalline cluster as a local order parameter.
Our simulation results indicate that the chain length only affects the nucleation rate but not the critical nucleus size, in agreement with previous theoretical and experimental work that formation of a critical nucleus is not influenced by the MW.
The nucleation rate broadly decreases with increasing chain length, and a further composition analysis of the critical nucleus suggests that intramolecular nucleation did not occur in our simulations.

We have demonstrated that relatively simple Monte Carlo simulations can be used with coarse-grained models of polymers to gain insight into their nucleation behaviour. Of course the polymer model we have used is particularly simple and it would be interesting to use a more realistic model to investigate real systems.
The interplay between initial nucleation, intra- and intermolecular growth and diffusion of polymer molecules and their individual segments makes for systems that exhibit particularly rich behaviours.
Hybrid Monte Carlo simulations within the framework of free-energy calculations enable us to gain insight into such systems without having to rely on complex collective Monte Carlo moves.
We hope that this preliminary investigation will stimulate further work on more realistic systems in the future.

\begin{acknowledgements}We acknowledge financial support from the National Key R\&D Program of China (2016YFB0302500), the National Natural Science Foundation of China (51633009), and the Royal Society Newton Mobility Grant MBAG/240 RG82754. We thank Prof.~Daan Frenkel and Prof.~Chuanfu Luo for fruitful discussions.
\end{acknowledgements}

Data supporting this study are openly available at the University of Cambridge Data Repository at \href{https://doi.org/10.17863/cam.51140}{doi:10.17863/cam.51140}.\footnote{\href{https://doi.org/10.17863/cam.51140}{doi:10.17863/cam.51140}}

\end{document}